\documentclass[twocolumn]{emulateapj}
\usepackage{graphicx}
\usepackage{natbib}
\begin{document}


\title{Exploring overstabilities in Saturn's A ring \\ using two stellar occultations.}
\author{M.M. Hedman$^{a,*}$, P.D. Nicholson$^{b}$, H. Salo$^c$}
\affil{\it  $^a$ Department of Physics, University of Idaho, Moscow ID 83844 USA \\
$^b$ Department of Astronomy, Cornell University, Ithaca NY 14853 USA \\
$^c$ Department of Physics, University of Oulu, Oulu Finland \\
${^*}$ Corresponding Author {\tt mhedman@uidaho.edu} \\
\\
{\em Proposed Running Head:} Overstabilities in Saturn's A ring. \\
\\}
\shorttitle{Overstabilities in Saturn's A ring.}

\begin{abstract}
 Certain regions of Saturn's rings exhibit periodic opacity variations with characteristic radial wavelengths  of up to a few hundred meters that have been attributed to viscous overstabilities. The Visual and Infrared Mapping Spectrometer (VIMS) onboard the Cassini spacecraft observed two stellar occultations of the star $\gamma$ Crucis that had sufficient resolution to discern a sub-set of these periodic patterns in a portion of the A ring between 124,000 and 125,000 km from Saturn center. These data reveal that the wavelengths and intensities of the patterns vary systematically across this region, but  that these parameters are not strictly determined by the ring's  average optical depth. Furthermore, our observations indicate that these opacity variations have an azimuthal coherence scale of around 3000 km. 
\end{abstract}

\maketitle

\section{Introduction}

Saturn's rings consist of many small particles in orbit around the planet. In denser ring regions, the gravitational and collisional interactions among the ring particles give rise to various fine-scale structures. In particular, certain parts of Saturn's rings exhibit periodic opacity variations with characteristic wavelengths between 100 and 500 meters. These patterns have opacity maxima that are nearly perfectly aligned with the azimuthal direction, and so  they act like diffraction gratings for radio signals passing through the rings, producing distinctive sidebands in Cassini Radio Science Subsystem occultations. Based on the distribution of these sidebands, \citet{Thomson07}  identified periodic structures in several parts of the main rings, including the  the inner A ring  (123,000-123,400 km and 123,600-124,600 km), the  inner B ring (92,100-92,600 km and 99,000-104,500 km) and the outer B ring (110,000-115,000 km). Extremely high-resolution opacity data obtained by the Cassini UVIS instrument during the occultation of the star $\alpha$ Leonis confirmed the existence of optical-depth variations with a characteristic wavelength of $\sim$160 meters in the outer B ring around 114,150 km \citep{Colwell07}.  Both of these observations indicated that these periodic variations are almost perfectly azimuthal.

\nocite{LO10}
These periodic structures have been interpreted as the result of viscous overstabilities in the rings \citep{Thomson07, Colwell07, Colwell09}.  Such overstabilities occur when the effective viscosity of the ring  increases sufficiently rapidly with increasing particle number density that oscillatory density variations can grow from small initial perturbations (see Schmidt {\it et al.} 2009 for a recent review of this phenomenon, with references to earlier work).  In hydrodynamical simulations without self-gravity, the characteristic wavelength of these  oscillations initially tends to increase with time \citep{ST95, Schmidt01, SS03}, but if the ring has a finite surface mass density, then non-linear phenomena and the ring's self-gravity limits the range of wavelengths that can be excited, leading to the  formation of highly periodic structures with wavelengths around  200 meters, similar to those observed in Saturn's rings (Schmit and Tscharnuter 1999, but see Latter and Ogilvie 2009, 2010 for other possible mechanisms for limiting the waves' growth). While fully self-gravitating N-body simulations have shown that overstabilities can exist in rings with finite mass densities and realistic inter-particle gravitational forces \citep{Salo01}, there is currently not a complete analytic theory for the formation of overstabilities in rings with finite mass densities, which may  simultaneously sustain non-radial fine-scale structure such as  self-gravity wakes  \citep{Schmidt09, LO09, RL13}. 

Two stellar occultations by the rings of the star $\gamma$ Crucis observed with the Visual and Infrared Mapping Spectrometer (VIMS) instrument onboard Cassini provide new information about the periodic structures in the inner A ring. The observations are described in Section~\ref{data}, while Section~\ref{wave} discusses trends in the wavelength and amplitude of the relevant patterns derived from a wavelet-based analysis of the light-curves. Section~\ref{turn} examines a subset of these data obtained when the star moved nearly azimuthally behind the rings in order to constrain the azimuthal coherence length of these patterns, which turns out to be $\sim10^{4}$ times the radial wavelength. Finally, Section~\ref{discussion} discusses some of the potential implications of these findings.

\section{Observations}
\label{data}

The stellar occultations discussed here were observed by the Visual and Infrared Mapping Spectrometer (VIMS) onboard the Cassini spacecraft. The VIMS instrument is described in detail in \citet{Brown04}, and is typically used to obtain spatially-resolved spectra of a scene. However, it can also operate in an ``occultation mode'' where the short-wavelength VIS channel is turned off, while the longer-wavelength IR channel stares at a single pixel targeted at a star and obtains a series of rapidly-sampled near-infrared stellar spectra at 31 wavelengths between 0.85 and 5.0 microns. Typically, we focus on the data from a spectral channel around 3 microns, where water ice is strongly absorbing and the rings are very dark. However, for this investigation we focus instead exclusively on data from one spectral channel covering the range 1.04-1.13 microns, because this channel has the best signal-to-noise (although the same patterns can be seen at other wavelengths). Even though the rings are not as dark at these shorter wavelengths, the occultations considered below involve a very bright star ($\gamma$ Crucis) passing behind the unlit side of the rings, so the sunlight scattered by the rings is still negligible compared to the flux from the star.

As the star moves behind the rings, its apparent brightness varies due to variations in the ring's opacity. The response of the detector is highly linear, so after a constant instrumental background is removed from each spectral channel, the data numbers (DN) returned by the instrument are proportional to the incident flux. Normalizing the stellar signal to its value outside the rings, we can translate the observed data numbers into estimates of the transmission through the rings $T$. These numbers can then be converted into the slant optical depth along the line of sight through the rings $\tau=-\ln(T)$, or the normal optical depth of the rings $\tau_n=\tau\sin |B|$ ($B$ being the elevation angle of the star above the rings), as needed.

The path of the star behind the rings is computed using a precise time stamp appended to each spectrum. These timing data, together with the appropriate reconstructed SPICE kernels (downloaded from the NAIF website\footnote{\tt http://naif.jpl.nasa.gov/naif/}), enable us to compute the radius and longitude at which the starlight pierced the ringplane for each sample in either occultation. This calculation accounts for the light travel time from the rings to Cassini, and uses stellar positions taken from the Hipparcos catalog\footnote{\tt http://heasarc.gsfc.nasa.gov/W3Browse/all/hipparcos.html}, corrected for proper motion and parallax at Saturn. Prior examinations of positions of sharp edges of gaps and ringlets in the Cassini Division and C ring in other occultations \citep{Nicholson11, French11} indicate that the reconstructed geometry is accurate to within a few hundred meters for radial cuts, most of which can be attributed to small uncertainties in the spacecraft's position along its orbital trajectory.

\begin{figure*}[tbp]
\centerline{\resizebox{6in}{!}{\includegraphics{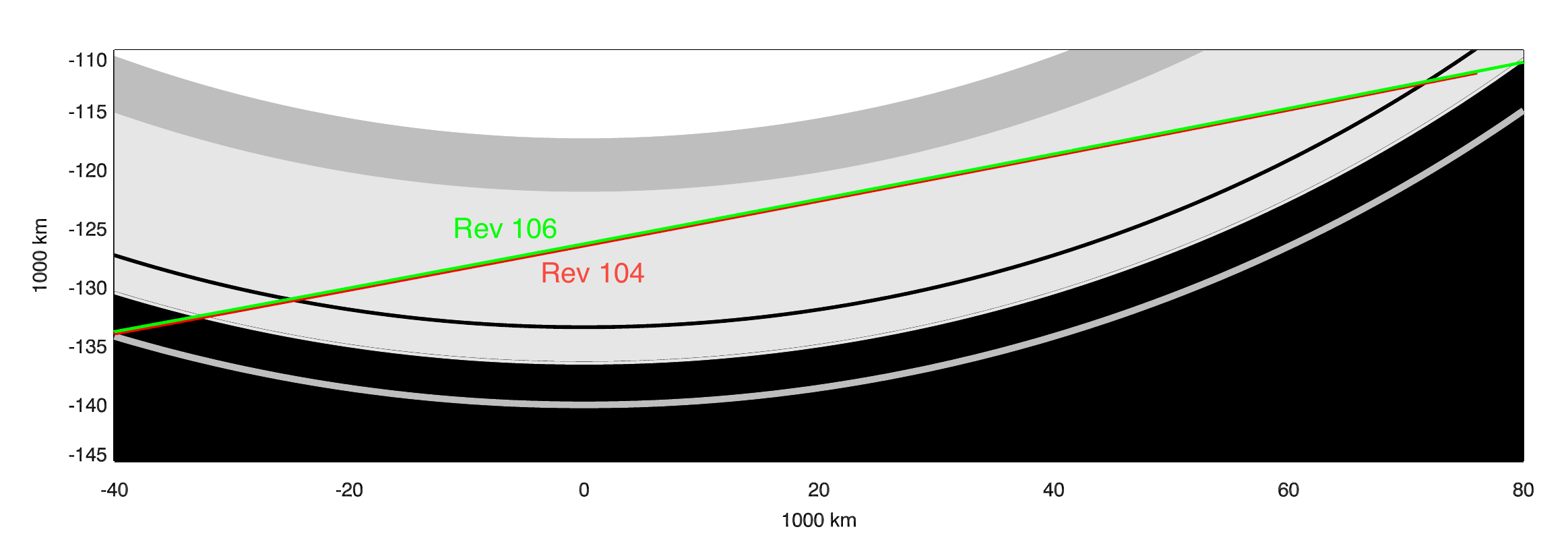}}}
\caption{Illustration of the Rev 104 and Rev 106 $\gamma$ Crucis occultation tracks behind Saturn's A ring. The background image shows the ring system from above, while the colored lines illustrate the apparent path of the star behind the rings during each occultation (Rev 104 in red and Rev 106 in green). In both cases the star moves from left to right. At first it moves towards smaller radii (``ingress"), but then turns around in the inner A ring and moves towards larger radii at the end of the occultation (``egress'').}
\label{tracks}
\end{figure*}

The occultations considered here involve $\gamma$ Crucis, a very bright star that is located well below the planet's ringplane ($B=-62.3^\circ$). Occultations by $\gamma$ Crucis therefore yield very good signal-to-noise ratios even in the more opaque regions where the periodic density variations are found.  More specifically, we will focus here exclusively on the occultations observed by Cassini on days 53 and 77 of 2009, which correspond to Cassini Revs\footnote{``Rev" is a designation for one of Cassini's orbits around Saturn} 104 and 106.  Figure~\ref{tracks} illustrates the track of the star behind the rings during these two occultations. Note these are not simple radial cuts through the rings but instead chords that reach a minimum radius, so each occultation can be divided into an ``ingress" and ``egress" component. The star reached a minimum radius of 124,413 km from Saturn center during the Rev 104 occultation and 124,262 km during the Rev 106 occultation. These ``turnaround''  radii  fall within one of the regions where sub-kilometer periodic structures were detected in radio occultations \citep{Thomson07}, and these particular occultation trajectories provide exceptionally fine radial sampling and resolution interior to 124,500 km.

Three factors can limit the resolution of a stellar occultation: (1) the star's finite size, (2) fresnel diffraction of the stellar signal and (3) the finite time it takes for the instrument to measure the star's brightness. Measurements performed with the VINCI instrument on the Very Large Telescope indicate that  $\gamma$ Crucis' angular diameter is  24.42$\pm$0.06 mas \citep{Kervella01}. This corresponds to a linear diameter of 130 meters when projected onto the ringplane for the two $\gamma$ Crucis occultations considered here. By contrast, the fresnel diffraction scale for these occultations (which depends on the light's wavelength and distance  between Cassini and the rings) is less than 50 m. Since the diffraction scale is so much smaller than the apparent stellar diameter, we can neglect diffraction for the purposes of this analysis. Finally, VIMS obtained a measure of the star's brightness every 20 ms for the Rev 104 occultation and every 60 ms for the Rev 106 occultation. Near the turnaround points, the occultation track is nearly azimuthal,  and the azimuthal speed of the stellar footprint relative to the orbiting ring particles is $\sim12$ km/s in the retrograde direction, so that successive samples are separated by $\sim 250$ m  in the Rev 104 occultation and $\sim 750$ m in the Rev 106 occultation. Thus the sampling time does limit our resolution in the azimuthal direction. However, the apparent radial velocity of the star is only $28*[(t-t_0)/100~s]$ m/s, where $t_0$ is the turnaround time, and the radial distance from the turnaround radius is $ \simeq 1.40*[(t-t_0)/100~s]^2$ km. Hence the radial sampling interval is less than a few meters within 6 km of the turnaround radius (or 200 sec from $t_0$). Indeed, the radial sampling interior to 124,500 km is so fine that the radial resolution in these regions is  limited by the apparent size of the stellar disk, which at 130 meters is small enough for opacity variations with wavelengths of order 200 meters to be clearly detectable.

\begin{figure}
\resizebox{3.5in}{!}{\includegraphics{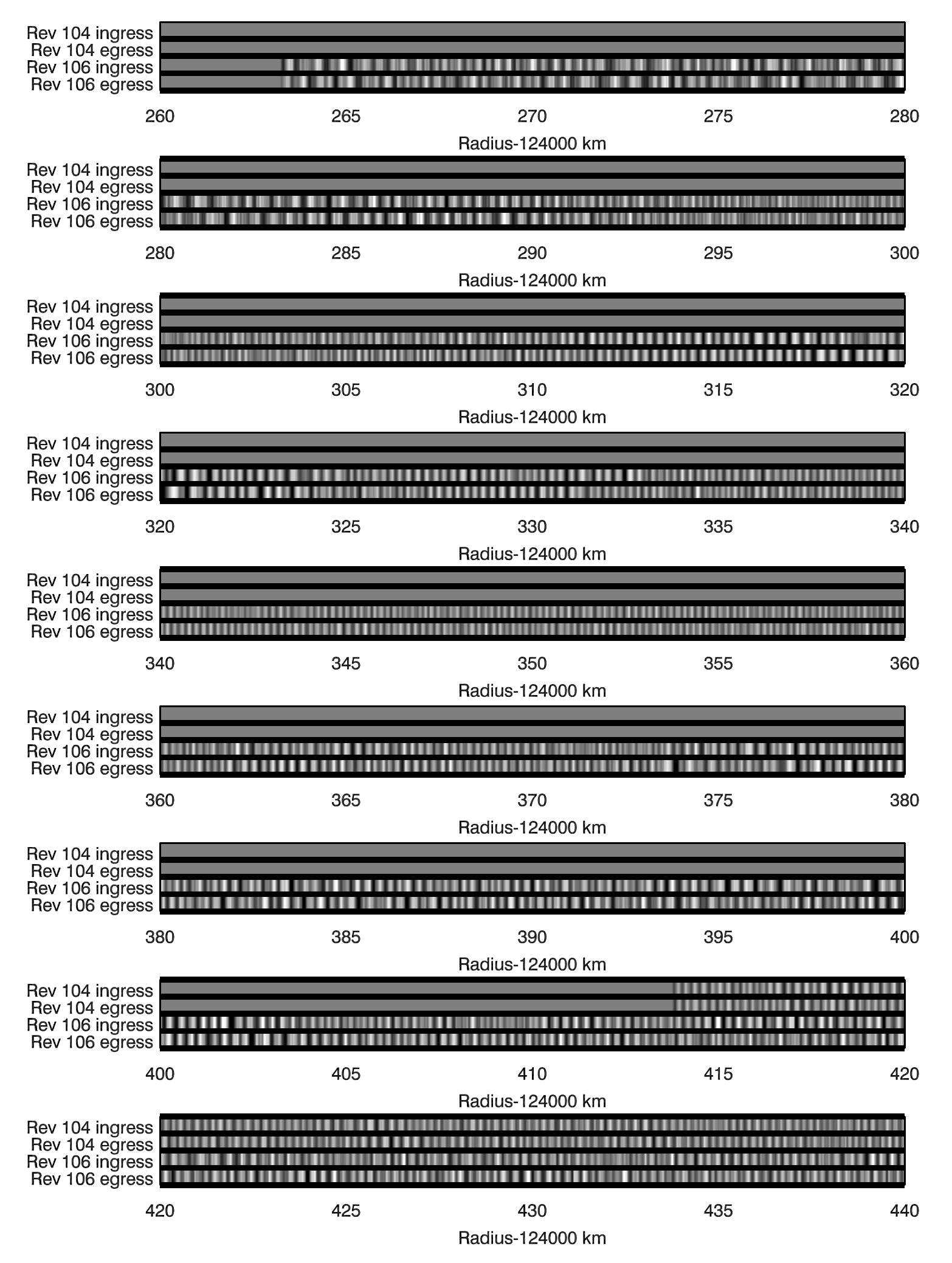}}
\caption{Illustration of the occultation data from the Rev 104 and Rev 106 $\gamma$ Crucis occultations. These two occultations yield a total of four cuts through the rings. Each profile is shown here as a greyscale image, with lighter shades corresponding to higher transmissions through the rings. These profiles have been filtered to remove signals on spatial scales smaller than 50 m and larger than 1 km, in order to highlight the periodic sub-kilometer patterns in these regions.}
\label{overview1}
\end{figure}

\begin{figure}
\resizebox{3.5in}{!}{\includegraphics{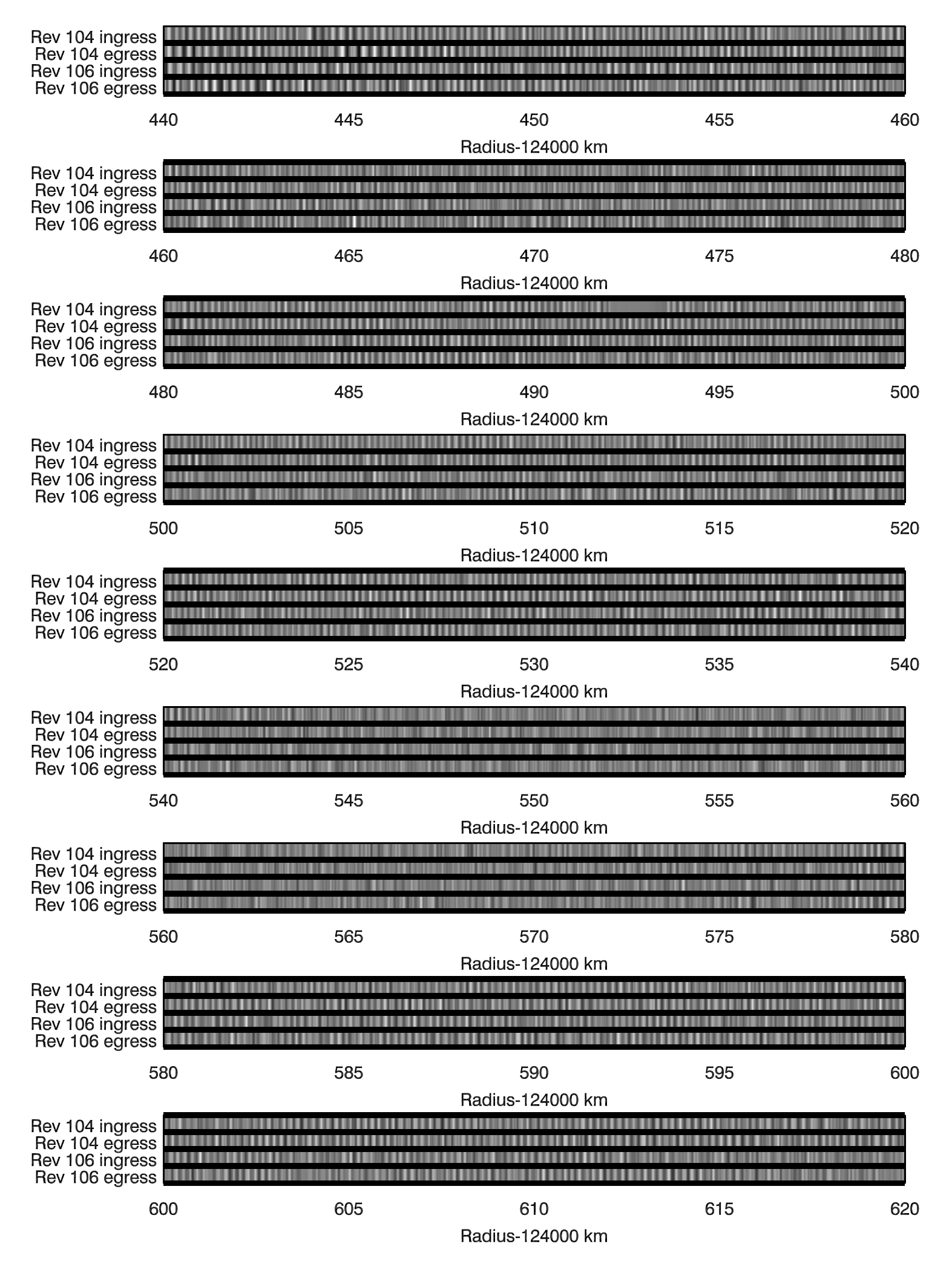}}
\caption{Illustration of the occultation data from the Rev 104 and Rev 106 $\gamma$ Crucis occultations. Each profile is shown here as a greyscale image, with lighter shades corresponding to higher transmissions through the rings. These profiles have been filtered to remove signals on spatial scales smaller than 50 m and larger than 1 km, in order to highlight the periodic sub-kilometer patterns in these regions. Note the abrupt disappearance of periodic brightness variations between 124,540 and 124,575 km}
\label{overview2}
\end{figure}

\begin{figure}
\resizebox{3.5in}{!}{\includegraphics{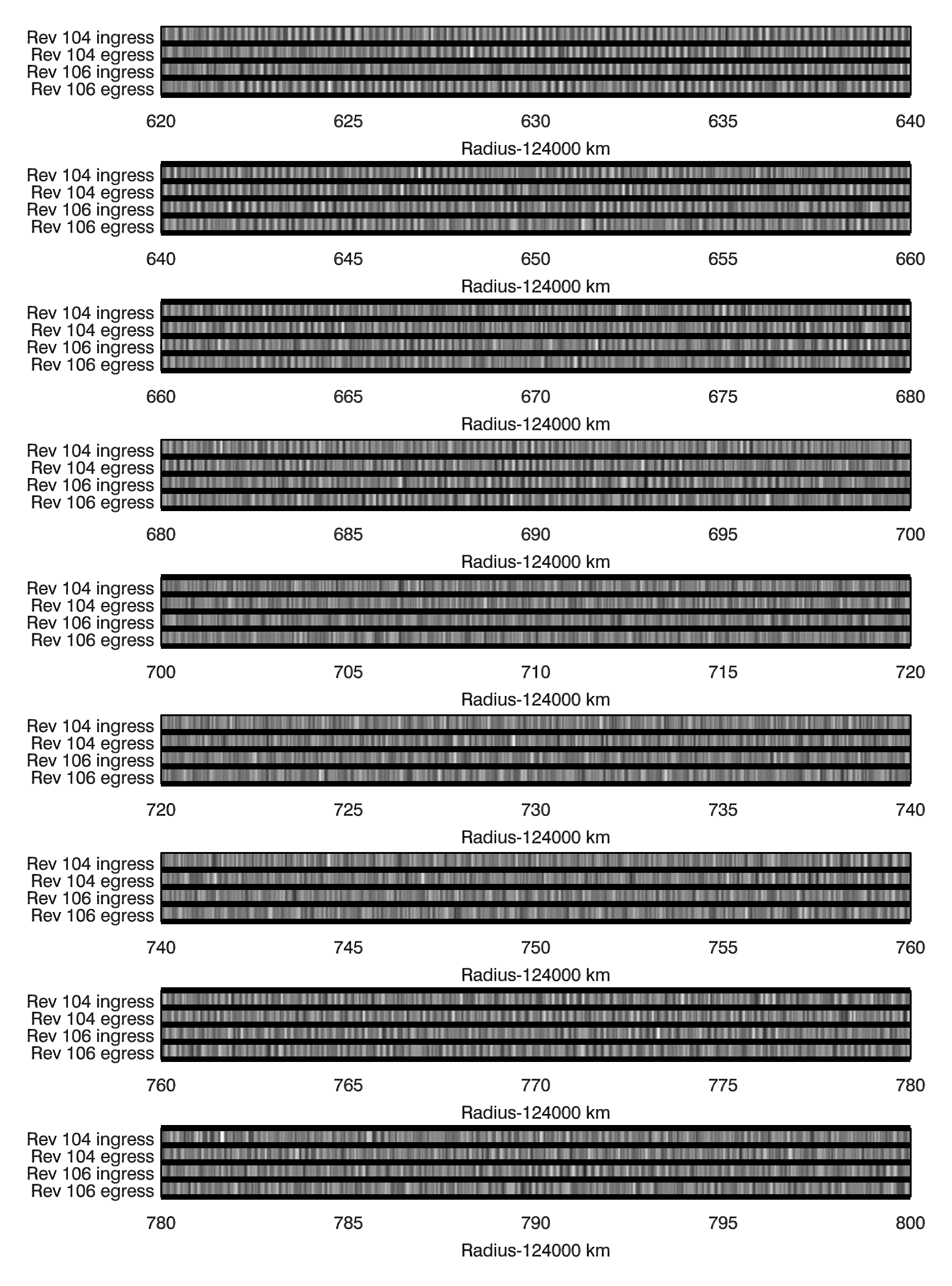}}
\caption{Illustration of the occultation data from the Rev 104 and Rev 106 $\gamma$ Crucis occultations.  Each profile is shown here as a greyscale image, with lighter shades corresponding to higher transmissions through the rings. These profiles have been filtered to remove signals on spatial scales smaller than 50 m and larger than 1 km, in order to highlight the periodic sub-kilometer patterns in these regions. Note the lack of clear periodic signals between 124,700 and 124,750 km.}
\label{overview3}
\end{figure}

 Figures~\ref{overview1}-\ref{overview3} illustrate the opacity profiles derived from these two occultations within  a few hundred kilometers of the turnaround points as greyscale images. Note that these images have been filtered to remove both broad-scale ($>1$km) radial trends in the optical depth and apparent fine-scale radial variations ($<50$ m) due to cosmic rays and stochastic azimuthal opacity variations, and thus emphasize the periodic sub-kilometer patterns, which appear as series of regular bright and dark bands.  These charts allow us to see the distribution of periodic signatures over fairly broad regions, and clearly show variations in the pattern's wavelength interior to 124,450 km, as well as some surprisingly well defined regions where the pattern seems to disappear (for example, between 124,545 and 124,575 km).  

\section{Amplitudes and wavelengths of the periodic patterns}
\label{wave}

In order to better understand these periodic patterns, we need to quantify how their amplitudes and wavelengths vary with position across the rings. These parameters are most efficiently extracted from the opacity profiles using a wavelet transform, which is essentially a localized Fourier transform that yields a measure of a periodic signal's strength and phase as a function of both spatial wavenumber and radial position \citep{TC98}. These tools have been used multiple times to investigate resonantly-driven waves in Saturn's rings \citep{Tiscareno07, Colwell09cd, Baillie11, HN13}.

\begin{figure*}
\centerline{\resizebox{6in}{!}{\includegraphics{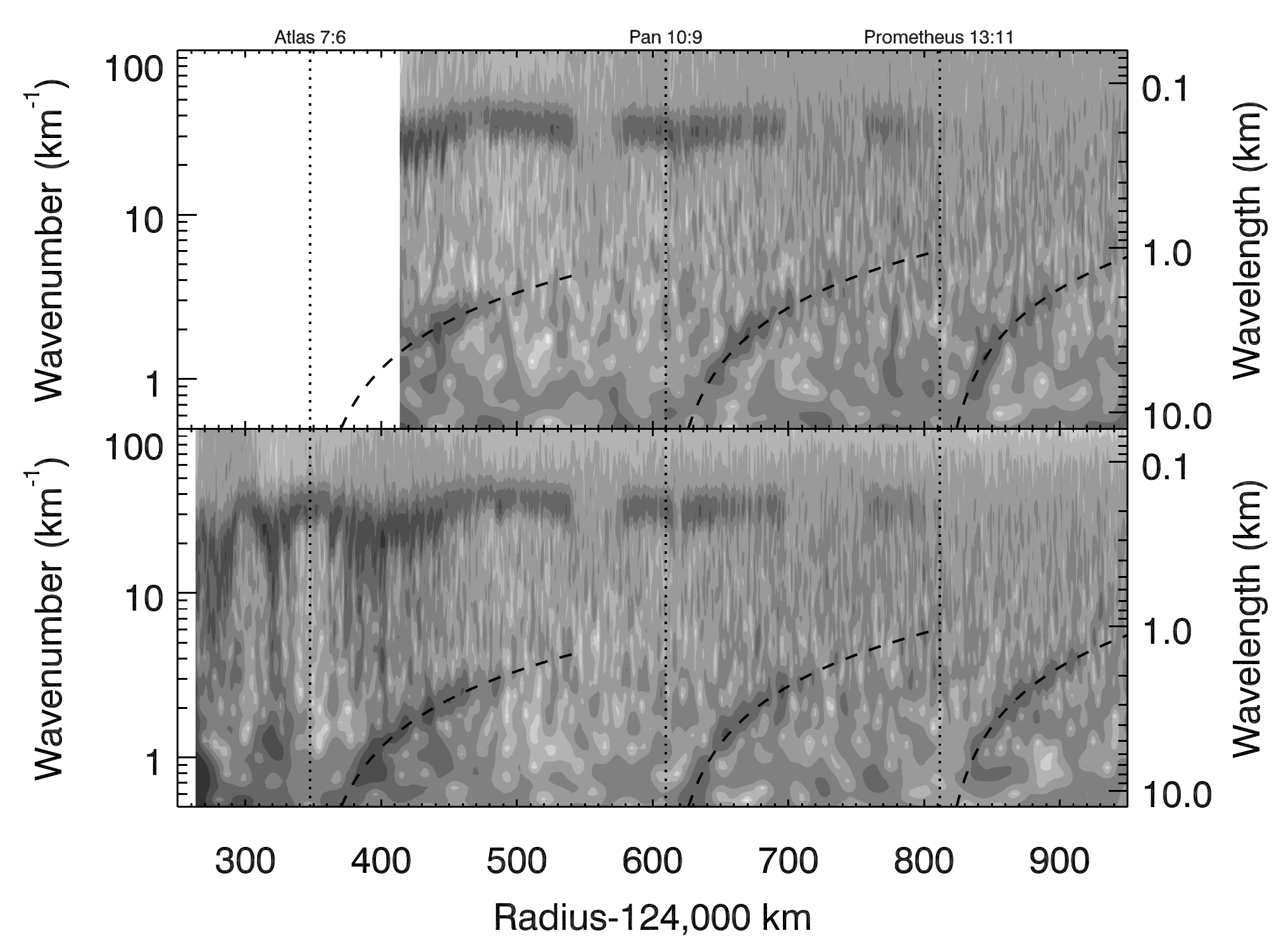}}}
\caption{Wavelet transform of the Rev 104 (above) and Rev 106 (below) occultation data. In both panels, shading indicates the signal strength as a function of ring radius and radial wavelength. Three weak density waves in this region produce signal on wavelengths longer than a kilometer, traced with the dashed lines. The periodic sub-kilometer patterns of interest here generate the bands near the top of each panel. Note that the radial resolution of these data has been reduced for display purposes.}
\label{waveletov}
\end{figure*}

Before we compute the wavelet transform for any of the $\gamma$ Crucis profiles, we must first interpolate each profile onto a regular array of radius values. For this analysis the profiles were interpolated and re-binned onto a regular array of radii with a sampling length of 20 meters prior to computing the wavelet.  This length was chosen to be well below the wavelength of the relevant patterns, but still large enough that the wavelet calculation could be performed in a reasonably short time. Figure~\ref{waveletov} illustrates the resulting wavelet transform of the occultation profiles computed using the pre-packaged IDL routine {\tt wavelet} (see Torrence and Compo 1998, the default Morlet mother wavelet with $\omega_0=6$ was used). These plots display the intensity of periodic signals as functions of both radius and wavelength, and the periodic sub-kilometer signal appears as a dark band running across the top of both panels at wavenumbers around 30 km$^{-1}$. Shifts in the wavelength of the pattern are seen as shifts in the vertical position of this band. This chart also shows significant signals at longer wavelengths associated with three relatively weak density waves launched from the Atlas 7:6, Pan 10:9 and Prometheus 13:11 inner Lindblad resonances.

\begin{figure*}
\centerline{\resizebox{6in}{!}{\includegraphics{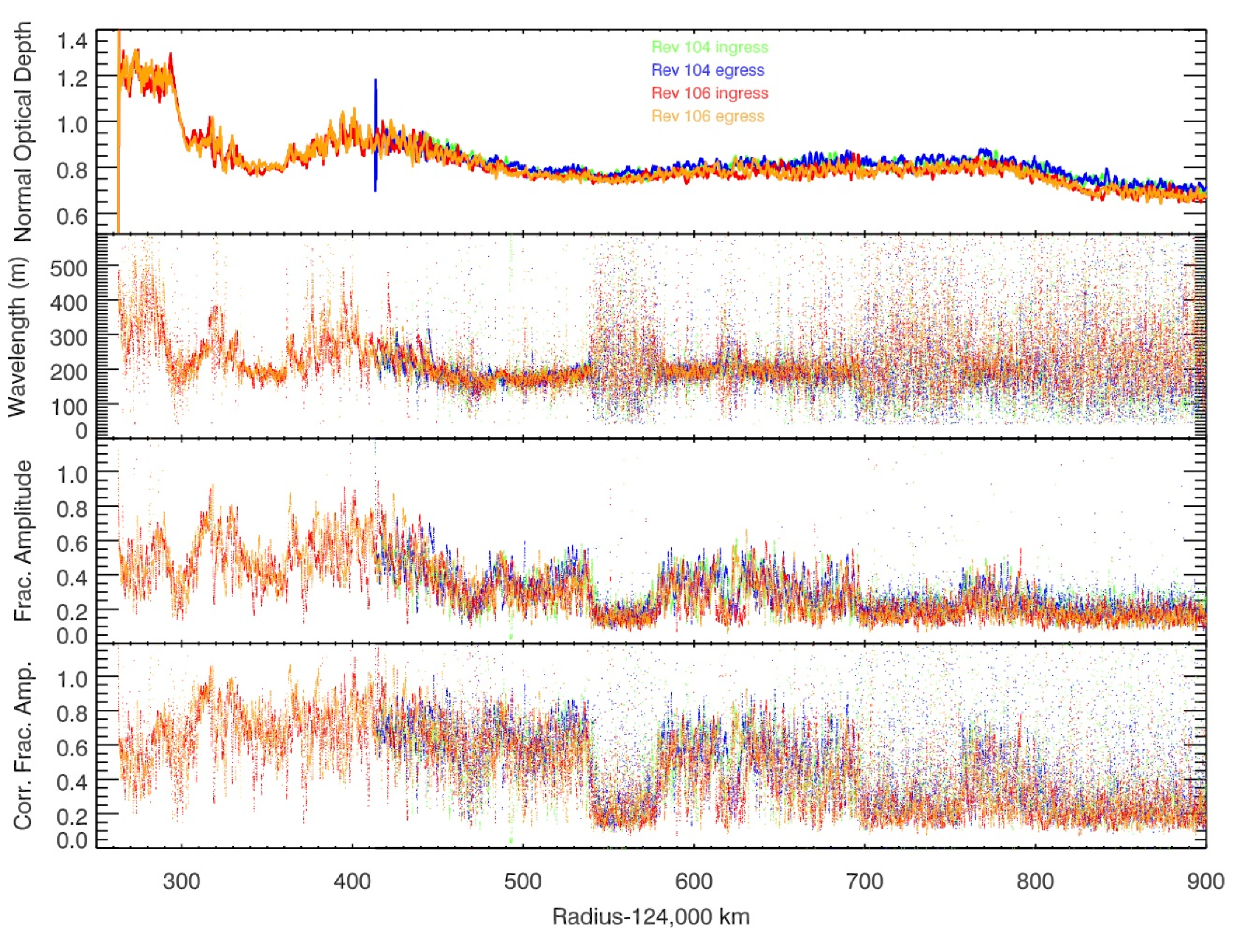}}}
\caption{Profiles of the sub-kilometer pattern's amplitude $A$ and wavelength $\lambda$ as a function of radius in the four occultation cuts. Each panel shows four profiles in different colors corresponding to the four different cuts through the ring. The top panel shows the average normal optical depth (smoothed to remove small-scale variations). The second panel shows the estimated wavelength of the sub-kilometer patterns. The third panel shows the observed fractional amplitude of the pattern, while the bottom panel shows the same amplitude with a correction applied to account for the star's finite angular extent (see text).}
\label{overprofs}
\end{figure*}

\begin{figure}
\resizebox{3in}{!}{\includegraphics{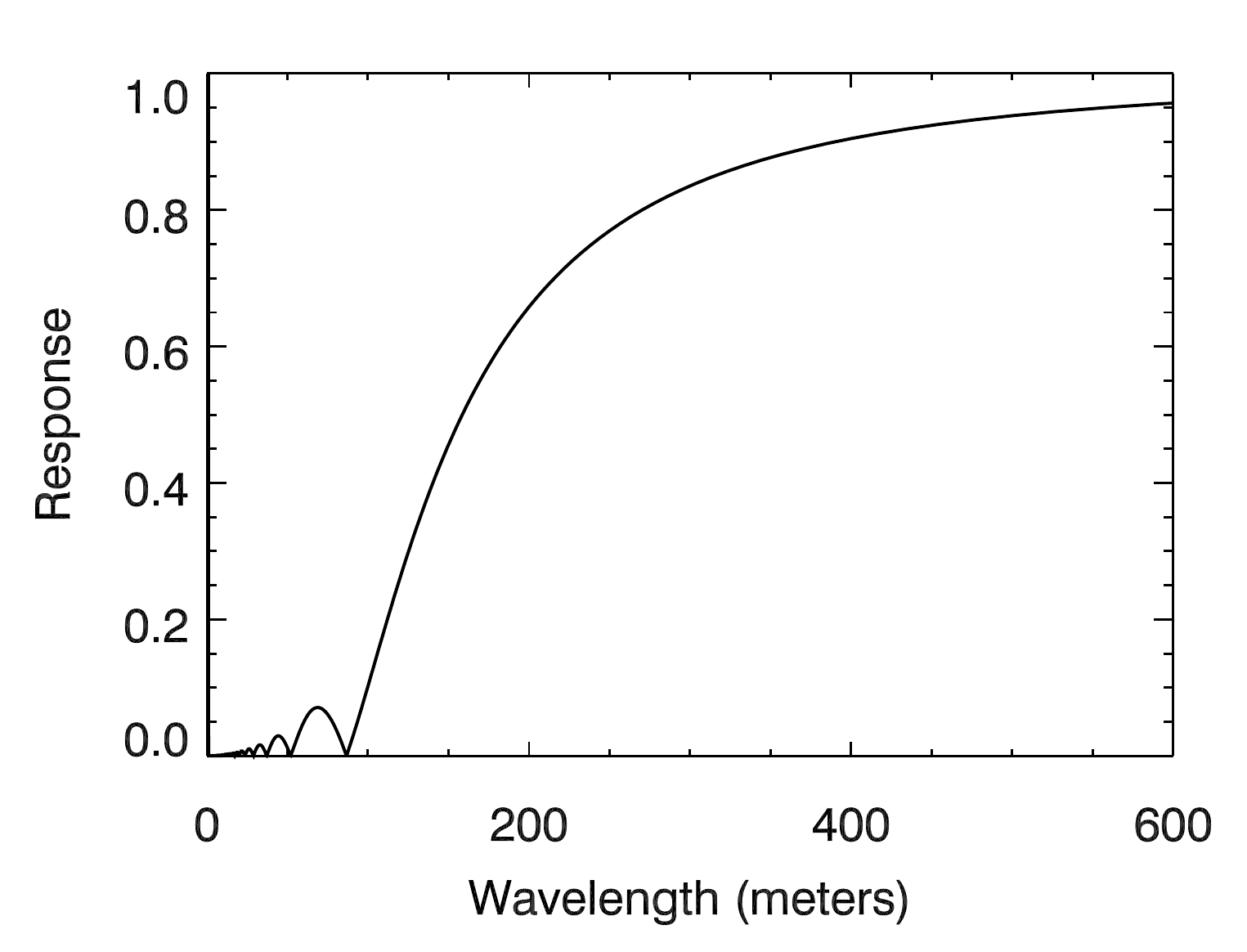}}
\caption{The filter function for a star with an apparent  diameter of 130 meters observed at normal incidence. The plot shows how much the amplitude of a sine-wave opacity variation is reduced when it is convolved with a uniform disk 130 meters across. As expected, this convolution reduces the amplitude of the measured signal significantly when the wavelength is about twice the stellar diameter.}
\label{starfilt}
\end{figure}

While standard wavelet routines such as this are sufficient to visualize the relevant periodic signals, more refined procedures are needed to  extract useful quantitative estimates of the pattern's wavelength $\lambda$ and fractional amplitude $A$. First of all, we computed the wavelet for the optical depth profiles rather than the raw transmission data in order to ensure that the measured fractional amplitudes will have sensible observational significance. Second,  we forced the relevant routines to compute the wavelet transform for a regular grid of wavelengths between 50 and 500 meters, sampled at an interval of one meter. This oversampling in the wavelength dimension facilitates the following calculations of the signal amplitude and wavelength, and the limited range of wavelengths considered automatically filters out longer-wavelength structures in the ring. To avoid confusion with the pattern's actual wavelength $\lambda$, the wavelength values used in the wavelet transform will be designated $\lambda_w$.

The wavelet transform of a given profile is a two-dimensional array of complex numbers as a function of radius and radial wavelength $\mathcal{W}(r, \lambda_w)$.  Let us denote the real and imaginary parts of the wavelet as $\mathcal{W}_R$ and $\mathcal{W}_I$, respectively. We can then define the wavelet power as $\mathcal{P}(r,\lambda_w)=\mathcal{W}_R^2+\mathcal{W}_I^2$ and the wavelet phase as $\varphi(r, \lambda_w)=\tan^{-1}(\mathcal{W}_I/ \mathcal{W}_R)$. Since we only compute the wavelet for a range of wavelengths between 50 and 500 meters, $\mathcal{W}$ only includes signals from the periodic sub-kilometer structures, and longer-wavelength signals from the density waves are automatically filtered out. 

At each radius $r$, the wavelet power exhibits a clear peak at a particular wavenumber. Provided the wavelet is properly normalized\footnote{We verified that for the wavelet transform used in this analysis, a sine wave of a given amplitude always produced a peak wavelet power equal to the amplitude squared.}, the magnitude of this peak can be directly transformed into an estimate of the pattern's (fractional) amplitude at this location:
\begin{equation}
A(r)=\frac{1}{\bar{\tau}(r)}\sqrt{max(\mathcal{P}(r,\lambda_w))},
\end{equation}
where $\bar{\tau}(r)$ is a smoothed version of the ring's optical depth profile (of course, the profile used in the wavelet is not smoothed). For this study, the smoothing was performed by applying a low-pass boxcar-averaging filter with an averaging length of 1 km in order to remove all the sub-kilometer variations from the normalization factor.

Ideally, the location of the peak wavelet power would provide an estimate of the pattern's wavelength. However, after some experimentation with wavelet transforms of simple sine curves, we found that more robust wavelength estimates can be derived from the wavelet phase\footnote{This result is consistent with previous work by \citet{Tiscareno07} and \citet{HN13}, both of which found the wavelet phase was the most useful way to quantify wavelengths in density wave profiles.}. At each radius, we derive an effective phase by first computing an effective average real part and imaginary part of the wavelet:
\begin{equation}
W_{R,I}(r)=\frac{\sum \mathcal{W}_{R,I}(r,\lambda_w)\mathcal{P}(r,\lambda_w)}
{\sum\mathcal{P}(r,\lambda_w)}
\end{equation}
where the sums are again over all values of $\lambda_w$. Note that these averages are weighted by the wavelet power, so that the wavenumbers with the strongest signals dominate the averages. From these average wavelet components, we can compute the average phase at each radius:
\begin{equation}
\phi(r)=\tan^{-1}({W}_I,{W}_R)
\end{equation}
By computing the average of components $\mathcal{W}_I$ and $\mathcal{W}_R$ instead of averaging the local phases $\varphi(r,\lambda)$, we avoid any difficulties involved in averaging a cyclic quantity. Note that this phase generally increases with increasing radius, and $\phi \simeq 0$ near any opacity maximum and $\phi \simeq \pi$ radians near any opacity minimum. Thus $\phi$ goes from 0 to $2\pi$ for each cycle of a periodic pattern, and we can estimate the pattern's dominant wavelength $\lambda$ at each radius from the radial derivative of the phase profile:
\begin{equation}
\lambda=2\pi\left[\frac{d\phi}{dr}\right]^{-1}.
\end{equation}

Figure~\ref{overprofs} plots profiles of the estimated pattern amplitudes and wavelengths as functions of radius, along with the average optical depth profiles for comparison. 
We may first note that there are clear variations in the wavelength and apparent amplitude interior to 124,500 km. Exterior to this, the pattern is either absent or has a wavelength of around 200 m, and the transitions between these two regimes appear rather abrupt, consistent with Figure~\ref{waveletov}. Finally, the pattern disappears completely exterior to 124,800 km. These results are reasonably consistent with the \citet{Thomson07} measurements of the periodic structures in these regions, which indicated that the average wavelength in this region is around 200 meters, with higher values closer to 124,300 km and lower values further from Saturn. However these observations indicate that the overstable patterns extend further out than \citet{Thomson07} suggested, and the reasons for this are still unclear (see Section~\ref{discussion} below).

However, we must take care when interpreting the apparent amplitude variations, because the observed amplitudes of these patterns are influenced by the finite size of $\gamma$ Crucis, whose angular extent corresponds to a projected diameter of about 130 meters at the ringplane (see Section 2).
This is not much less than the wavelengths of these patterns. The star's finite size therefore acts as a low-pass filter on the occultation data, causing the pattern's amplitude to be underestimated by a wavelength-dependent factor. Indeed, there appears to be a strong positive correlation between the pattern's wavelength and its observed amplitude in Figure~\ref{overprofs}.

If we neglect limb darkening and assume the observed line-of-sight is exactly perpendicular to the ring-plane, then the radial structure in the ring will be convolved with a function of the form:
\begin{equation}
S(r-r_o)=\left\{
\begin{array}{c l}     
    \cos\left(\frac{r-r_o}{\pi D_*}\right) & |r-r_o|<D_*/2\\
    0 & |r-r_o|>D_*/2
\end{array}\right.
\end{equation}
where $D_*$ is the star's projected diameter. Figure~\ref{starfilt} shows the filter function associated with this convolution. As expected, the amplitude reduction becomes more and more pronounced as the wavelength decreases, with the signal being almost entirely suppressed when the wavelength is less than $D_*$. 

We can use this function to correct the observed amplitudes and estimate the true amplitude of the opacity variations in this region. These corrected amplitudes are shown in the bottom panel of Figure~\ref{overprofs}. We must caution that these calculations do not account for the star's limb darkening, if any,  and do not include projection effects that arise from the starlight passing through the ring at an incidence angle of $27.7^\circ$. Even so, it is interesting to note that after this first-order correction, the opacity variations are bimodal, being either near the noise floor of the observations ($A<0.2$) or having fractional amplitudes of around 0.7, which implies that the density variations are either absent or close to saturation throughout this region. There may be a slight reduction in the fractional amplitudes interior to 124,300 km, which could be associated with the rapid increase in normal optical depth in this region, but even here the fractional opacity variations  exceed 0.5. This is consistent with these patterns being generated by an overstability, which will cause density/opacity variations to grow until they saturate due to non-linear processes \citep{ST99, Salo01, SS03, Schmidt09, RL13}.

By comparison, the estimated wavelength of these signals are much less sensitive to the stellar diameter than their amplitudes are. Convolutions are products in Fourier space, and the typical widths of the peaks in the wavelet power $\mathcal{P}$ are only around 10-15\% the peak wavelength, and so the filter function cannot shift the estimated peak location as easily as it can influence the peak amplitude. Indeed, some basic numerical experiments show that the filter function will only produce a small ($<$10\%) bias in the estimated wavelengths to longer values. Hence the wavelength trends shown in Figure~\ref{overprofs} are most likely real. As mentioned in the introduction, there is not yet a full analytical theory that would predict how the wavelength should vary depending on the rings' mass density, viscosity, etc. Thus the trends identified here could potentially provide some guidance in this regard. Between 124,350 and 124,550 km, the pattern wavelength seems to be correlated with the mean optical depth, with both reaching a broad maximum around 123,400 km. However, we should point out that this is also the region occupied by the Atlas 7:6 density wave, so both the ring's optical depth and its pattern wavelength might be reacting to that perturbation. Furthermore, interior to 124,350 km, the correlations between optical depth and pattern wavelength are rather weak. While the high optical-depth region interior to 124,300 km does exhibit a somewhat higher  wavelength than other regions, comparably long wavelengths are found at  124,320 km, which is at substantially lower optical depth. Also, in between these two regions, around 124,300 km, there is a pronounced minimum in the wavelength, where the optical depth shows a sharp gradient between two levels. This suggests that the wavelength of these patterns depends not only on the local optical depth, but  also on other parameters, such as the opacity or density gradient. 

We may also note that no periodic patterns are detectable exterior to 124,820 km, where the normal optical depth falls below 0.8. This suggests that a normal optical depth of around 0.8 is the minimum necessary for these patterns to exist. However,  there are some regions with optical depths between 0.8 and 1.0 where a periodic pattern is clear and other regions with the same range of optical depths where no periodic patterns can be detected. Thus the existence of such patterns, like their wavelengths, does not appear to be solely a function of opacity.  

\section{Azimuthal coherence of the patterns}
\label{turn}

\begin{figure}
\resizebox{3.5in}{!}{\includegraphics{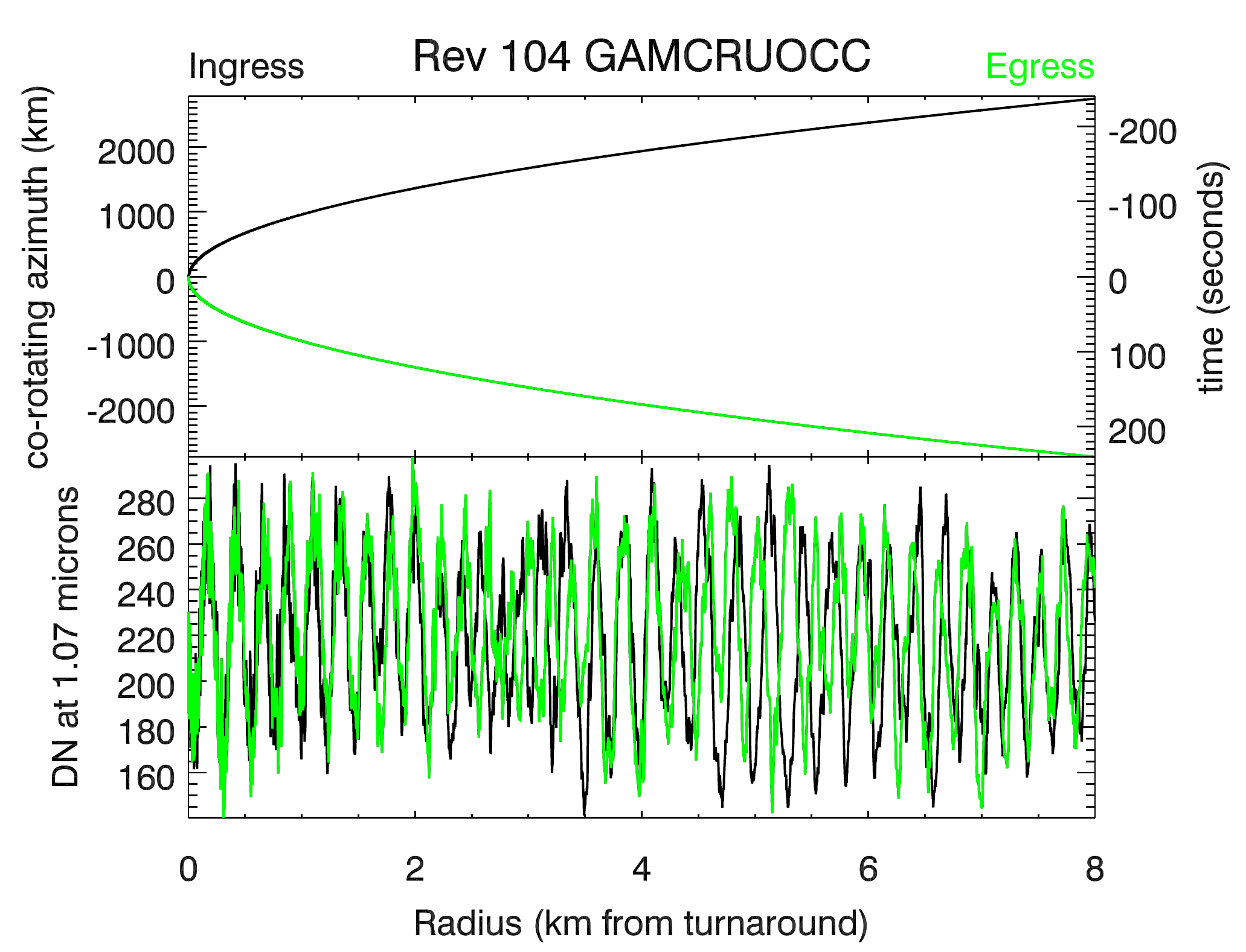}}
\caption{Plot showing the data around the turnaround point for the Rev 104 $\gamma$ Crucis occultation. The top panel shows the co-rotating azimuth and time elapsed as a function of radial distance from the turnaround point (at 124,413.26 km), with the ingress and egress legs colored black and green respectively.  The bottom panel shows the data number (DN) profile (smoothed over 20 samples for clarity) on the same radial scale. Note that interior to 2 km, the ingress and egress profiles are almost perfectly aligned, but further from the turnaround the two patterns go in and out of step.}
\label{radturn4}
\end{figure}

\begin{figure}
\resizebox{3.5in}{!}{\includegraphics{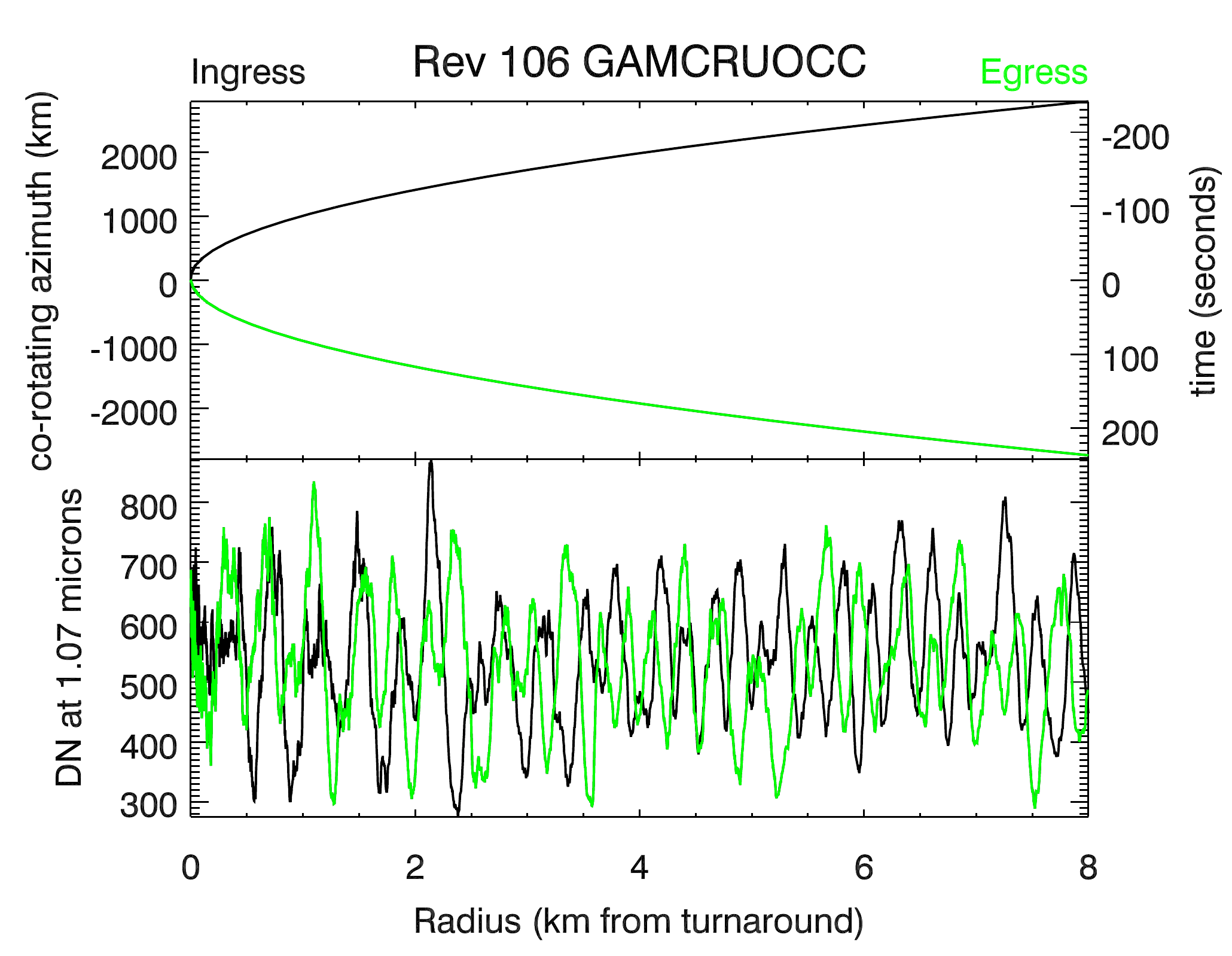}}
\caption{Plot showing the data around the turnaround point for the Rev 106 $\gamma$ Crucis occultation. The top panel shows the co-rotating azimuth and time elapsed as a function of radial distance from the turnaround point (at 124,262.75 km), with the ingress and egress legs colored black and green respectively.  The bottom panel shows the data number (DN) profile (smoothed over 20 samples for clarity) on the same radial scale. The correlation between the ingress and egress profiles near the turnaround is not nearly as strong as in the Rev 104 data, but peaks and troughs generally align within 2 km of the turnaround radius, with a noticeably lower correlation further out.}
\label{radturn6}
\end{figure}

\begin{figure}
\resizebox{3.5in}{!}{\includegraphics{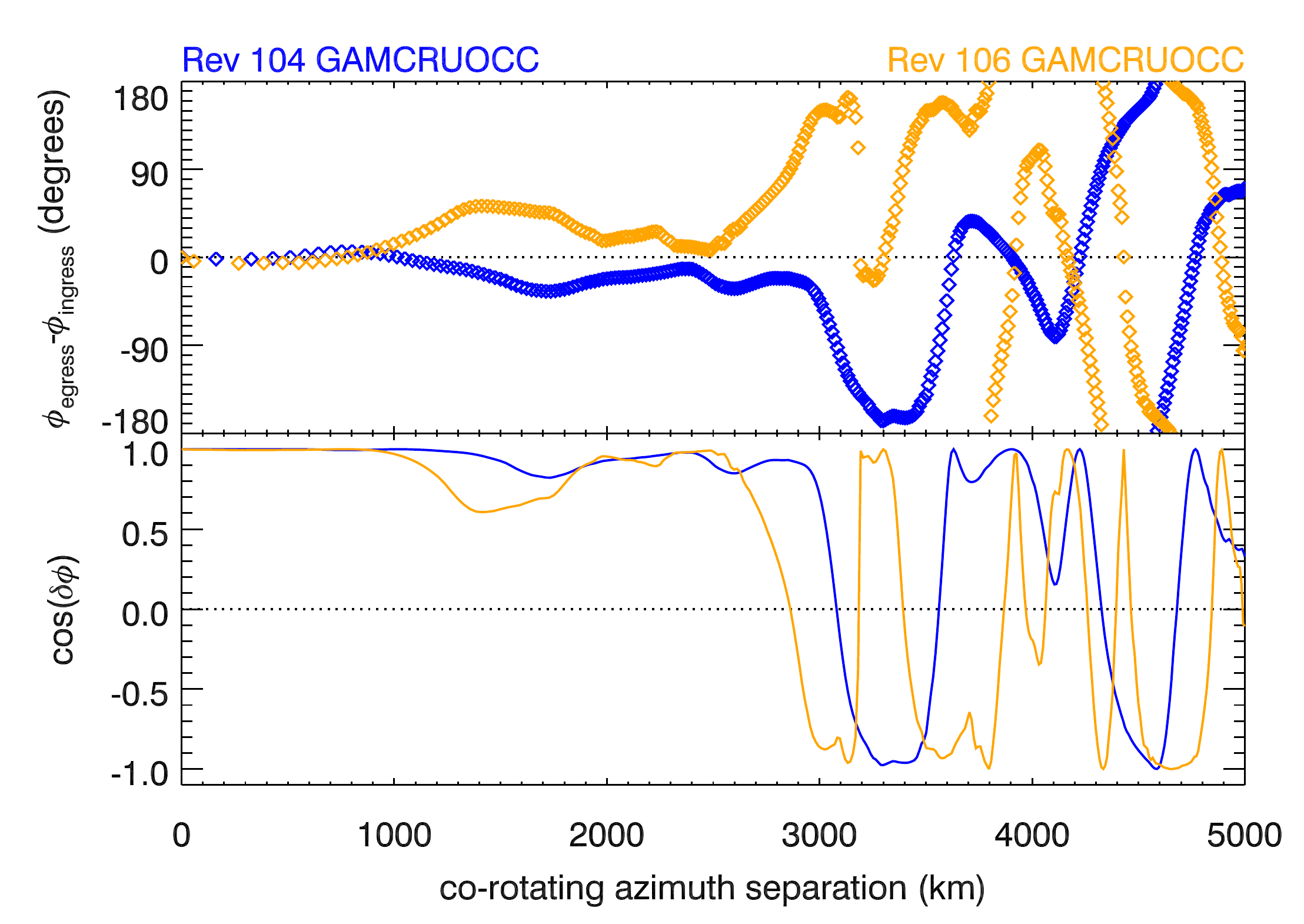}}
\caption{Plots showing the wavelet phase difference between the ingress and egress cuts for each occultation as a function of co-rotating azimuth separation between the two cuts. Note that for both the Rev 104 and Rev 106 data the phase differences are rather low, and the correlation coefficient $\cos\delta \phi$ is above 0.5 out to around 2500 km. However, by around 3000 km, the phase differences begin to wander over all possible values. This suggests that the typical azimuthal coherence lengths of these overstable patterns is of order 3000 km.}
\label{dphase}
\end{figure}

Additional insights into the nature of these patterns can be obtained by examining the occultation data near the turnaround points, where the star was moving almost purely in the azimuthal direction. Figures~\ref{radturn4} and~\ref{radturn6} illustrate both the geometry and the light curves from these innermost portions of the occultations. The upper panels on these plots show the observed radius as a function of both time and co-rotating azimuth $r(\theta-nt)$, where $\theta$ and $t$ are the observed inertial longitude and time measured relative to their values at the turnaround point, and $n$ is the calculated orbital mean motion at the observed radius $r$. Note that the observed co-rotating azimuth decreases with time because the ring material was overtaking the star as it moved slowly across the ring.

For both occultations, the star spent less than 10 minutes within 8 km of the minimum track radius. This is much less than the 12.5-hour local orbital period at these locations, which is also the expected oscillation period of any overstable pattern \citep{Salo01, RL13}. Thus we may reasonably assume that the pattern did not evolve significantly over the course of these observations. However, during this time, the relative motion of the ring and star causes the track to cover over 4000 km in co-rotating azimuth, which provides an unprecedented opportunity to explore the azimuthal coherence scale of these patterns. All  numerical simulations of overstabilities conducted thus far have had comparable radial and azimuthal dimensions, and in all these simulations the pattern appears to be purely azimuthal. However, the simulations apply periodic boundary conditions in the azimuthal direction which will certainly favor azimuthal symmetry. In reality, the radial wavelengths of these patterns are  less than a kilometer, so we would not expect them necessarily to remain coherent for thousands of kilometers, much less all the way around the ring, but just how far a given pattern might extend is still unclear. These data allow us to finally quantify how azimuthally symmetric these patterns really are.

Looking at the Rev 104 data first, we find the ingress and egress cuts match up remarkably well within 2 km of the turnaround point, which indicates that the pattern is coherent and axisymmetric over scales of up to 2500 km. However, further from the turnaround point, the two patterns begin to shift relative to each other, and are sometimes exactly in phase and sometimes 180$^\circ$ out of phase. This suggests that the azimuthal coherence scale of the pattern is not much greater than a few thousand kilometers in this portion of the A ring.

Turning to the Rev 106 data, we find that the correlation between the ingress and egress profiles is not nearly so good, despite the fact that the pattern's wavelength is noticeably longer in this region. This suggests that the azimuthal coherence in this region (124,262 km) is somewhat less than in that probed by the Rev 104 occultation. However, within 2 km of the turnaround, we still find that peaks and troughs are aligned, indicating that there is some correlation between the structure in these two cuts. However, further out, we again see that even this poor correlation falls apart, with  all possible phase shifts being found between the ingress and egress cuts. Intriguingly, this breakdown in the correlation again appears to occur just beyond 2 km from the turnaround, which suggests that the azimuthal correlation length in this region is also of order a few thousand kilometers.

A more quantitative way to compare the two profiles is to examine the difference in the wavelet phase parameter $\phi$ between ingress and egress. If the two profiles are well aligned, then this difference $\delta \phi$ should remain close to zero, but if the two patterns are completely uncorrelated, $\delta\phi$ will cycle around all possible values. Indeed, the correlation coefficient between the two profiles at a given location should simply be $\cos\delta\phi$. Figure~\ref{dphase} plots both $\delta\phi$ and $\cos\delta\phi$ as a function of the co-moving azimuth separation between the two cuts. Consistent with the profiles in Figure~\ref{radturn4}, the $\delta \phi$ values for the Rev 104 data remain close to zero out to about 2500 km, or 2.5 km from the minimum radius, then things slip a little, come back into in phase at 3000 km, and finally one profile appears to skip one cycle at an azimuthal separation of $\sim$4500 km (i.e. 4-6 km from the  turnaround radius).  The Rev 106 data show a similar pattern, although the phase differences close to the turnaround point are somewhat larger on average than they are in the Rev 104 data. These plots suggest that in both regions the azimuthal coherence length of the patterns is of order 3000 km, or 0.025 radians/1.5$^\circ$ in azimuth. 

There are a number of different ways that the patterns could become de-correlated between the ingress and egress cuts.  One possibility is that there are ``domains'' of purely azimuthal patterns separated by well-defined boundaries. While we have no clear idea what  such a domain boundary would look like, we do not find any evidence in either profile for the pattern disappearing or  the phase difference changing very abruptly,  and so we do not have direct evidence to support this particular scenario. Another option is that the patterns become de-correlated more gradually because the density maxima are not perfectly aligned with the azimuthal direction, but instead drift back and forth in radius. In this scenario, the observed coherence length indicates that the local cant angle of the density crests is of order $10^{-4}$ radians (i.e. 300m/3000 km). 

\section{Discussion}
\label{discussion}

Currently, the best explanation available for these periodic, almost purely azimuthal sub-kilometer patterns is that they represent viscous overstabilities. However, additional theoretical and observational work is needed to validate or refute this interpretation. While established linear theories can provide information about the conditions required to initiate overstability, the non-linear processes responsible for determining the final amplitude and wavelength of the patterns are still not perfectly understood, especially for situations where the mutual self-gravity of the particles cannot be ignored \citep{ST99, Salo01, Schmidt09, LO09, RL13}. Since we cannot compare our observations with detailed theoretical predictions for the behavior and distribution of overstabilities, we will instead discuss briefly two ways these measurements could help inform future theoretical work on overstable phenomena. On the one hand, this new information about the radial distribution of periodic patterns could help guide numerical simulations geared towards determining the parameters that influence the expression of overstabilities. On the other hand, our constraints on the patterns' azimuthal coherence lengths will likely require the development of novel theoretical analyses before they can be properly interpreted.

\begin{figure*}[tbp]
\centerline{\resizebox{6.5in}{!}{\includegraphics{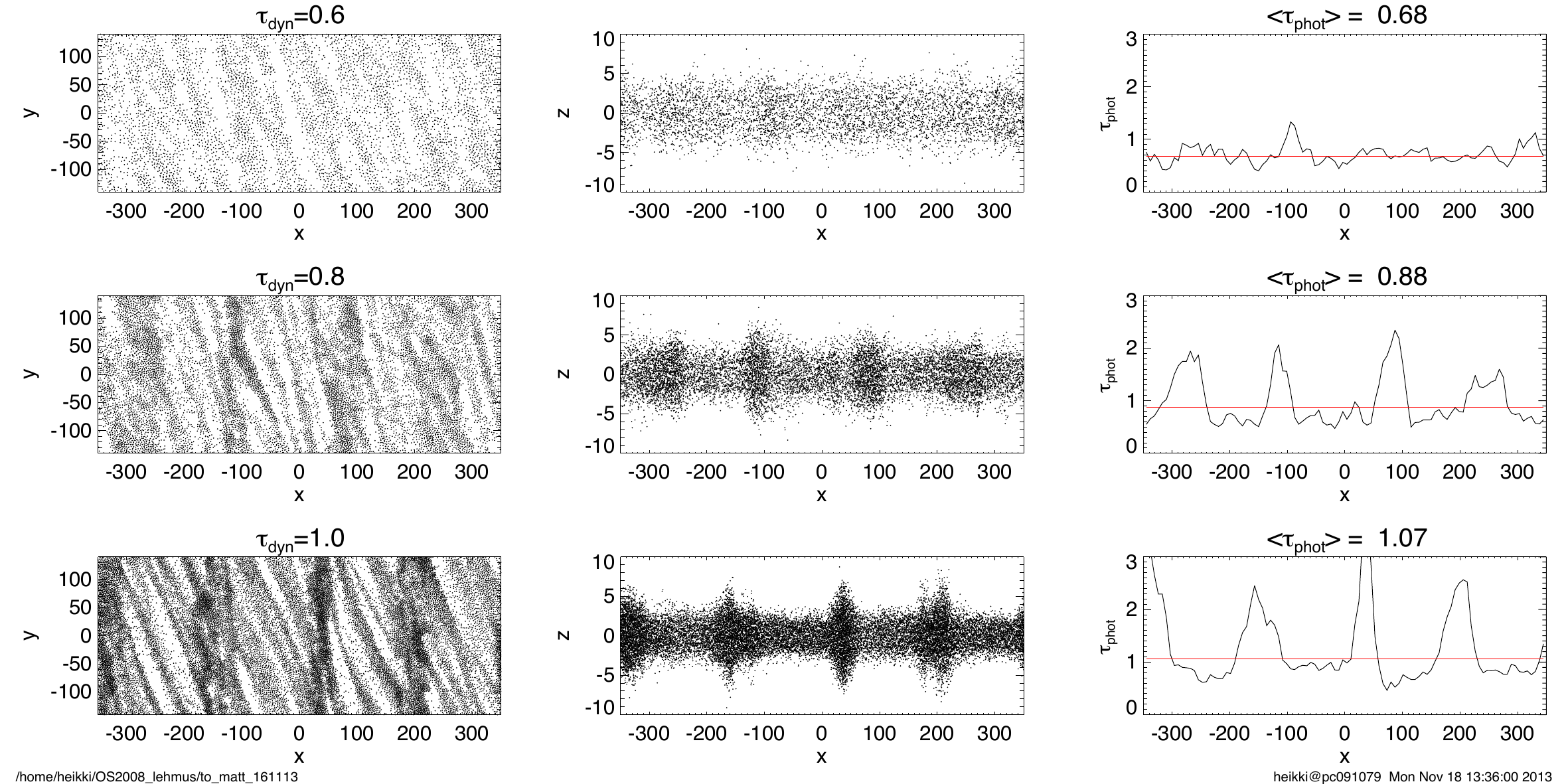}}}
\caption{Sample simulations of small ring patches (700 m$\times$280 m) around 124,000 km with dynamical optical depths of 0.6, 0.8 and 1.0 (corresponding to average photometric optical depths of .68, .88 and 1.07).  Each simulation consists of identical particles with coefficient of restitution $\epsilon = 0.1$, particle internal density of 225 kg/m$^3$ and an average surface density of 500 kg/m$^3$. The left panels show snapshots of the simulations after 200 orbit periods viewed from above ($x$=radial direction, $y$=azimuthal direction) when the system has reached a steady state. The middle panels show the same simulation in a side (or edge-on) view, while the right plot shows the average photometric optical depth versus radius. While no overstable pattern is visible in the $\tau=0.6$ case, clear overstable density variations are visible in both higher opacity cases.}
\label{simfig}
\end{figure*}

Most numerical simulations of dense rings consider a small patch of ring material because of limitations on available computer time. Provided the area of the patch is larger than the wavelength of the relevant overstable patterns, such simulations are sufficient  to ascertain the conditions under which overstabilities are likely to develop. For example, Figure~\ref{simfig} shows the results of three numerical simulations of a ring with three different optical depths, computed using the algorithms described in \citet{Salo95} and ~\citet{Salo01}, which include the mutual gravitational interactions among the relevant ring particles. These patches all have the same surface mass densities but different ``dynamical optical depths'' $\tau_{dyn}$ (a standardized measure of the total cross sectional area of all particles in the simulation) of 0.6, 0.8 and 1.0, which correspond to average observable normal optical depths of 0.68, 0.88 and 1.07, respectively. All three simulations exhibit an array of tilted features known as self-gravity wakes, which arise from the competition between the particles' mutual gravitational attraction and Keplerian shear. Since the surface density is the same in all three simulations, the radial separation of wake structures (set by Toomre's critical wavelength) is the same. However, in two of the simulations there are also  periodic axisymmetric patterns that corresponds to a viscous overstability.  These simulations show that there is a relatively sharp transition between optical-depth regimes  where the overstability does not occur (e.g. the $\tau_{dyn}$=0.6 case)  to those where the opacity variations associated with the overstability are substantial (e.g. the $\tau_{dyn}$=0.8 case). This seems to be consistent with our observations that show rather sharp transitions between regions where the overstability is clear and where it is undetectable (see also Rein \& Latter 2013).  

For these particular simulations, overstabilities develop when the observed optical depth is around 0.8, which is close to the minimum optical depth where periodic patterns can be detected in the VIMS observations. However, these models assume that the ring particles are all the same size and have a rather low particle internal density of 225 g/cm$^3$, so more work needs to be done to ascertain whether the same threshold can be achieved with higher particle densities and different particle sizes, surface densities, etc. Also, the VIMS observations show that the pattern's distribution across the rings is not strictly a function of the ring's optical depth. For example, no periodic pattern is visible around 124,570 km and 124,720 km, even though the latter region has an optical depth that exceeds 0.8 and appears to be high enough to support patterns elsewhere in this part of the rings. Even if we limit our attention to regions where we can clearly detect a periodic structure, the pattern's wavelength is not always correlated with optical depth. For instance, we observe a minimum in the pattern's wavelength around 124,300 km, where the optical depth undergoes a sharp transition from 1.2 to 0.8. It remains unclear whether these trends can be explained by shifts in the local particle size distribution or other particle properties, or if they require consideration of other aspects of the dynamical environment. Probably the best way to address these issues is by numerically simulating a wide array of conditions in order to thoroughly explore the relevant parameter space.

While currently available numerical tools should be able to clarify the radial distribution of the periodic patterns, understanding the azimuthal properties of these structures will likely require different theoretical approaches. Simulating ring regions hundreds or thousands of kilometers across is beyond the abilities of current techniques, so it may be some time before theorists can simulate a region large enough for the overstable regions to exhibit real non-axisymmetric features. While smaller simulations could potentially reveal how localized structures like embedded moonlets could disrupt the overstable patterns,  semi-analytical models of how non-circular density ridges evolve over time (analogous to those used to describe density waves and other resonant structures) may also yield useful insights. 

In lieu of such theoretical advances, we can use the observational data to try and better understand how the patterns' coherence scale depends upon its dynamical environment. Both $\gamma$ Crucis occultations indicate that the azimuthal coherence length of the periodic patterns in the inner A ring is  substantial, of order 3000 km. However, such long coherence scales may not be a universal feature of these periodic patterns. For example, UVIS detected periodic structures with a typical wavelength of 160 meters in an occultation with a turnaround radius around 114,150 km in the outer B ring \citep{Colwell07}. Unlike the VIMS occultations, the ingress and egress profiles derived from these UVIS observations are not obviously correlated with each other close to the turnaround point. This suggests the patterns observed by UVIS have a much shorter azimuthal coherence length, perhaps less than 100 km. This reduced coherence length might reflect the particular environment probed by the UVIS occultation. The outer B ring exhibits strong stochastic optical depth variations on a broad range of scales which make the opacity profiles obtained during different occultations very poorly correlated, and the weak correlation between the ingress and egress profiles in the UVIS data could be just another expression of this general behavior. However, it is also possible that the weak correlations in the UVIS data are due to the much higher optical depth in the outer B ring ($\tau_n \sim 3$)  compared to the inner A ring ($\tau_n \sim 1$). One potential bit of supporting evidence for this second option is that the periodic patterns seen near the turnaround in the Rev 106 VIMS occultation, where $\tau_n \sim 1.2$, are less well correlated than the patterns seen in the Rev 104 data, where $\tau_n \sim 0.9$. Thus it might be that increasing the opacity of the rings tends to reduce the azimuthal coherence length of the overstable patterns. Probably the best way to confirm or refute this idea would be with additional data from other occultations that turn around at different locations within regions possessing periodic patterns. 

Another potentially useful tool for understanding the coherence lengths of these patterns are the sidebands \citet{Thomson07} described in the radio occultation data. Since these sidebands occur where a periodic opacity variation acts like a diffraction grating for the radio signal, the intensity of these sidebands depends both on the magnitude of the density variations and their coherence lengths. It is interesting to note that the radio science experiment only detected sidebands out to around 124,400 km, while VIMS detected periodic signals out to 124,800 km. This difference could be due to time-variability in the extent of the periodic patterns between 2005 and 2009, but may also indicate that the patterns beyond 124,400 km are not coherent enough over a large enough region to produce proper sidebands in the radio occultation. If the latter is the case, the radio occultations could provide useful information about the coherence scales of these periodic patterns and how they vary across the rings.

\section*{Acknowledgements}

We wish to thank the VIMS team, the Cassini Project and NASA for generating the data used in this analysis. This work was supported by the Cassini Data Analysis Program Grant NNX11AN92G.


\end{document}